\documentclass[prl, showpacs, preprint]{revtex4}
\usepackage{graphicx,epsfig}
\usepackage{bm}
\usepackage{dcolumn}
\usepackage{color}
\begin{document}
\title{Mapping the dynamics of complex multi-dimensional systems onto a discrete set 
of states conserving mean first passage times: a Projective Dynamics approach}
\author{Katja Sch\"afer}
\email{ks379@msstate.edu}
\affiliation{Department of Physics and Astronomy, Mississippi State University, Mississippi State, MS 39762-5167}
\author{M.~A.~Novotny}
\affiliation{Department of Physics and Astronomy, Mississippi State University, Mississippi State, MS 39762-5167}

\begin{abstract}
We consider any dynamical system that starts from a given ensemble of configurations 
and evolves in time until the system reaches a certain fixed stopping criterion, 
with the mean first-passage time the quantity of interest.  
We present a general method, Projective Dynamics, 
which maps the multi-dimensional dynamics of the system onto an 
arbitrary discrete set of states $\left\{\zeta_k\right\}$, 
subject only to the constraint that the dynamics is restricted to transitions not further 
than the neighboring states $\zeta_{k\pm 1}$.  
We prove that with this imposed condition there exists a master equation with 
nearest-neighbor coupling with the same mean first-passage time as the original dynamical system.  
We show applications of the method for Brownian motion of particles
in one and two dimensional potential energy landscapes and the folding process of small bio-polymers. 
We compare results for the mean first passage time and the mean folding time obtained with the 
Projective Dynamics method with those obtained by a direct measurement, and where possible with a 
semi-analytical solution.
\end{abstract}

\maketitle
\section{Introduction}

The mean first-passage time (MFPT) is the time required for an ensemble of identical 
systems starting from a given initial distribution of configurations 
to reach a fixed criterion.  Being able to obtain the MFPT is important in a wide array of physical 
applications.  These include the MFPT when a star of a certain mass goes supernova, 
to neuron firing dynamics, spreading of diseases, 
chemical kinetics, folding processes of polymers, the death of living creatures, 
and the decay of metastable states via nucleation and growth processes \cite{redner,cond07}.  
The Projective Dynamics (PD) method to calculate MFPT 
for systems with a discrete state space \cite{prl1,Kol03,slwf,proc1} mapped the MFPT problem 
for metastable decay of magnetic states associated with Ising-type (discrete spin states) 
models onto a nearest-neighbor coupled master equation.  However, the PD method has not previously been 
shown to be valid for other dynamical systems, for example for systems with continuous 
state spaces \cite{me}.  The calculation of the MFPT for systems with 
long time scales has previously required that the 
dynamic system satisfy certain conditions.  For example, for the milestoning method \cite{elber} 
to work the system must be close to equilibrium or have separation of time scales or committer surfaces.  
We prove in this letter that the PD method has no such constraints, and hence it is generally 
applicable to obtain the MFPT of {\it any\/} dynamical system.  

We prove that for any MFPT problem (with a finite MFPT) there exists a discrete 
master equation with nearest-neighbor coupling that has the same MFPT as the original system.  
This master equation has time-independent rates, even if the underlying system is non-Markovian.  
Every point in phase space in the original system must be mapped onto a state in the discrete 
master equation.  We show that {\it any\/} mapping that leads to only nearest-neighbor coupling 
in the master equation preserves the MFPT provided the growing ($g_k$) and shrinking ($s_k$) 
rates between all master equation states $k$ are correct.  
Therefore, whether the $g_k$ and $s_k$ are known theoretically or are measured, the correct MFPT 
is obtained.  For metastable decay in the ferromagnetic $d$$=$$3$ Ising case 
such knowledge enabled calculations of MFPT of about $10^{50}$ 
 Monte Carlo steps per spin \cite{Kol03}.  
It is anticipated that the knowledge that the PD method gives the correct MFPT will enable 
long-time-scale simulations for other physical models.   An example of where the PD method 
may be useful for long-time simulations is studies of the intercalation time of Li ions 
in molecular dynamics studies modeling charging of batteries \cite{AbouHamad}.  

\section{Theory}
Consider any ensemble of systems evolving in phase space 
$\Gamma$ under some dynamic.  They start with some given distribution, and 
we desire to calculate the MFPT until each system stops after it meets some 
fixed stopping criterion.  The dynamics of the ensemble can be mapped 
onto a set of discrete non-overlapping states $\left\{\zeta_k\right\}$, with 
$0$$\le$$k$$\le$$S$, which cover the 
entire domain.  We impose the restriction that the states $\zeta_k$ can make transitions 
only to the states $\zeta_{k\pm 1}$.  
This condition can be met for many systems by 
adjusting the `length' $\Delta \zeta$ of the states, so that no transitions within a 
single time step $\mbox{d}t$ further than
to the neighboring states exist.  The states that form the stopping citerion 
will all be mapped to $\zeta_0$.
 
The time-evolution of the system can be described by a master
equation with nearest neighbor coupling, which describes the change of occupation 
$P_k(t)$ of state $k$  by the 
probability flows between adjacent states. Namely,
\begin{equation}
\frac{\mbox{d}}{\mbox{d} t}P_k(t)=g_{k+1}P_{k+1}(t)-\left[ g_{k}+s_{k}\right] 
P_k(t)+s_{k-1}P_{k-1}(t) \mbox{,}
\label{master1}\end{equation}
where the growing $g_k$ (shrinking $s_k$) rates represent the directional rates at 
which the system transits
from state $k$ to state $k-1$ ($k+1$). 
The initial conditions $\{ P_k(0)\}$ 
are specified by the ensemble of initial conditions of the original system.

Let $N(k\rightarrow k\pm 1;t,t+\mbox{d}t)$ denote the number of times the system makes
a transition from state $k$ to its neighboring states ($k \pm 1$) in the 
time interval $(t,t+\mbox{d}t]$ and $N_k(t)$
the number of times the system resides in state $k$ at time $t$. Then the growing rate
for the time interval $(t_o,t_e]$ can be obtained by measuring 
the system at time intervals $\mbox{d}t$. 
Explicitly  
\begin{equation}
g_k \text{d} t = \frac{\int_{t_o}^{t_e}\mbox{d}tN(k\rightarrow k-1;t,t+\mbox{d}t)}
{\int_{t_o}^{t_e}\mbox{d}t N_k(t)} \>.
\label{grodef}
\end{equation}
A similar expression exists for the shrinking rate $s_k$.  

Assuming that the ensemble of systems has been fully absorbed at 
$t_e=\infty$ after starting at $t_o=0$, 
we write in shorthand the master equation 
\begin{equation}
\frac{\mbox{d}}{\mbox{d}t}P_k(t)=\sum_{k'} W(k,k') P_{k'}(t) \>,
\end{equation}   
where the elements of $W(k,k')$ are the growing and shrinking rates, which have been measured 
from $t_o=0$ up to
some point in time when all systems in the ensemble are fully absorbed.
By using the fundamental relation \cite{zwanzig}
$
\sum_{k}\tau(k)W(k,k')=-1,
$
the general expression for the MFPT $\tau(S_o)$ for a system 
starting fully in state $S_o$ 
can be obtained as
\begin{equation}
\tau (S_o)=\sum_{i=1}^{S_o}\frac{1}{g_i}+\sum_{\ell=1}^{S-1}\left[ \sum_{i=1}^{\min (S_o,|S-\ell|)}
\left( \frac{\prod_{j=i}^{i+\ell-1} s_j}{\prod_{j=i}^{i+\ell}g_j}\right) \right]
\label{001}
\end{equation}
where $S_o$ denotes the starting state and $S$ the total number of states within the domain.

We prove that Eq.~(\ref{001}) gives the same MFPT independent of the particular choice of the 
states. For this purpose, let
$m\equiv k \bigcup k+1$ be the joined interval of state $k$ and $k+1$, then for a single time step 
it follows that the transitions from state $m$ to $k-1$ ($k+2$) in the new system are 
equal to the transitions
from state $k$ to $k-1$ ($k+1$ to $k+2$) in the old system 
namely 
$\tilde{N}(m \rightarrow k-1,t,t+\mbox{d}t) = N(k \rightarrow k-1,t,t+\mbox{d}t)$ and 
$\tilde{N}(m \rightarrow k+2,t,t+\mbox{d}t) = N(k+1 \rightarrow k+2,t,t+\mbox{d}t)$, 
whereas the probability
of occupation of state $m$ rescales as $\tilde{P}_m(t)=P_k(t)+P_{k+1}(t)=[N_{k}(t)+N_{k+1}(t)]/\sum_k N_k(t)$. 
Thus the growing rate at time $t$ in the old and new system are related as
\begin{equation}
\tilde{g}_m
= \frac{P_k(t)}{P_k(t) + P_{k+1}(t)} g_k
\label{gres} 
\end{equation}
and similarly for the new shrinking rate ${\tilde s}_m$.

That this procedure leaves the MFPT unchanged can be understood by noting 
the effect of joining two states on the master equation. The joining of adjacent states 
corresponds
to the addition of terms of the master equation of the same adjacent states, namely 
\begin{equation}
\frac{\mbox{d}}{\mbox{d} t}P_k(t)=g_{k+1}P_{k+1}(t)-\left[ g_{k}+s_{k}\right] 
P_k(t)+s_{k-1}P_{k-1}(t) \mbox{.}
\label{master2}\end{equation}
and
\begin{equation}
\frac{\mbox{d}}{\mbox{d}t}P_{k+1}(t)=g_{k+2}P_{k+2}(t)-\left[ g_{k+1}+s_{k+1}\right] 
P_{k+1}(t)+s_{k}P_k(t) \mbox{.}
\label{master3}\end{equation}
It is easily seen that the cross terms vanish and we are left with an expression of the form
\begin{equation}
\frac{\mbox{d}}{\mbox{d}t}\tilde{P}_m(t)=g_{k+2}P_{k+2}(t)- g_{k} P_{k}(t)-s_{k+1} 
P_{k+1}(t)+s_{k-1}P_{k-1}(t) \mbox{.}
\label{masterend}\end{equation} 
Using (\ref{gres}) and 
similarly for ${\tilde s}_m$ 
in (\ref{masterend}) for a single time step, leads to 
\begin{equation}
\frac{\mbox{d}}{\mbox{d}t}\tilde{P}_m(t)=g_{k+2}P_{k+2}(t)- [\tilde{g}_{m}+\tilde{s}_{m}] 
\tilde{P}_{m}(t)+s_{k-1}P_{k-1}(t) \mbox{.}
\end{equation} 
Hence the general form [compare Eq.~(\ref{master1}) and (\ref{master2})] 
is restored in the reduced system of equations. 
Performing this joining repeatedly, finally only one state is left. This gives the 
same MFPT as the original ensemble of systems \cite{slwf}.  

\begin{figure}[ht]
    \includegraphics[width=\linewidth]{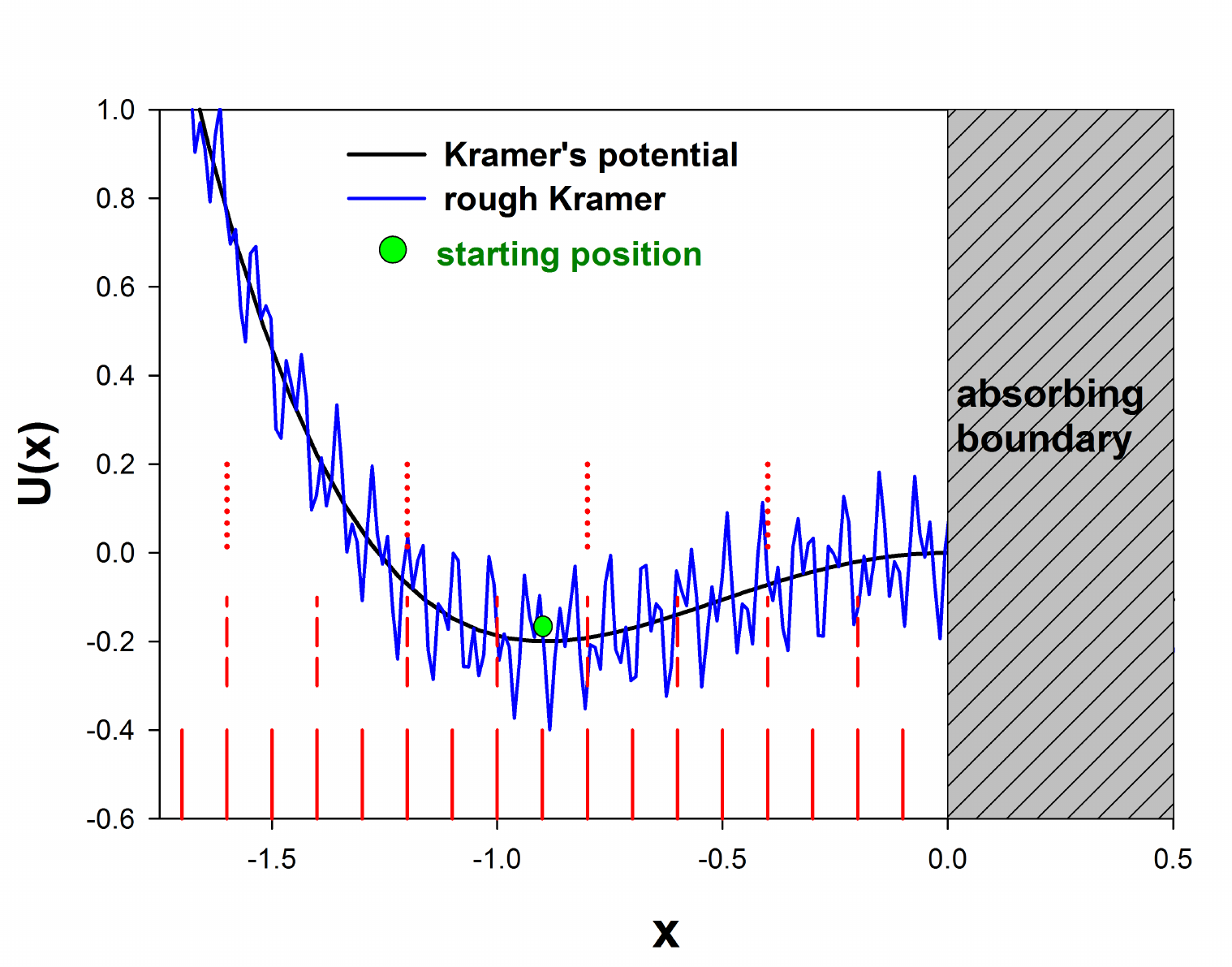}
\caption[]{(color online) Kramer's and a rough potential. The states are chosen as intervals 
along the $x$-coordinate of some 
fixed length $\Delta l$.}
\label{kramer}
\end{figure}

\section{Computational Study/Simulation}
To illustrate the theoretical framework, we demonstrate diffusion in both dimensions 
$d$$=$$1$ and $d$$=$$2$, as well as the folding process of a model for a small linear polymer chain.  
We choose these examples to specifically address the 
independence of the method from any 
existence or knowledge of a free energy landscape or most probable escape pathway.

As a $d$$=$$1$ historical example, we studied the diffusion process of 
particles subjected 
to a Kramers or to a rough \cite{Zwan88} potential (Fig.~\ref{kramer}). The trajectories 
were integrated 
along 5120 individual paths starting at $x$$=$$-$$0.9$ and terminated once they leave the 
domain $(-\infty,0]$ at the absorbing boundary $x=0$. The MFPTs 
were calculated using Eq.~(\ref{001}). In order to show the 
freedom in choosing the length of the intervals, we show 
examples of bins with different lengths $\Delta l$. 
Starting at the absorbing boundary $k$$=$$0$ for $x$$\ge$$0$, 
the system resides in state k 
when $(k-1)\Delta l \le x<k\Delta l$. In both cases (smooth and rough), we chose 
$\Delta l = 0.1,0.2 
~ \text{and}~ 0.4$ [Fig.~(\ref{kramer})].  
Table~\ref{Tab2a} shows 
the results obtained with the PD method are not changed with different 
interval lengths. A comparison with a direct measurement 
(on the underlying dynamics) and the semi-analytical solution 
further verifies that the MFPTs obtained with the PD 
method gives reliable 
results (provided that sufficient statistics are obtained to get accurate estimates of 
the growing and shrinking rates). 
Note that to illustrate the method we chose examples of the starting position 
being at the boundary between two 
states ($\Delta l=0.1$), in the midrange ($\Delta l=0.2$), and away from the midrange 
($\Delta l=0.4$) of the starting state. 
The results do not change with any of these starting positions, 
which further indicates the independence of the PD method.

\begingroup
\squeezetable
\begin{table*}
\renewcommand{\arraystretch}{1.4}
\setlength\tabcolsep{5pt}
\begin{tabular}{l|llll|ll}
\hline\noalign{\smallskip}
$k_{\text{B}}T$ & Projective & Dynamics\\
\hline\hline
~ & ~ & ~ & ~&~ & Brownian & Riemann\\
~ & $\Delta l=0.1$ & $\Delta l=0.2$ & $\Delta l=0.4$ &~& motion & integration\\
\hline
$0.1^a$ & $18.4(\pm 0.2)$ & $18.4(\pm 0.2)$ & $18.4(\pm 0.2)$ & ~&$18.4(\pm 0.2)$ & $18.44$  \\
$0.0875^a$ & $24.7(\pm0.3)$ & $24.7(\pm0.3)$ & $24.7(\pm0.3)$ &~& $24.7(\pm0.3)$ & $24.80$ \\
$0.075^a$ & $36.3(\pm 0.4)$ & $36.3(\pm 0.4)$ & $36.3(\pm 0.4)$ &~& $36.3(\pm 0.5)$ & $36.46$ \\
$0.0625^a$ & $62(\pm 1)$ & $62(\pm 1)$ & $62(\pm 1)$ &~& $62(\pm 1)$ & $61.9$\\
\hline
$0.1^b$ & $46.3(\pm 0.6)$ & $46.3(\pm 0.6)$ & $46.3(\pm 0.6)$ &~& $46.3(\pm 0.6)$ & 46.2  \\
$0.0875^b$ & $80 (\pm 1)$ & $80 (\pm 1)$ & $80 (\pm 1)$ &~& $80 (\pm 1)$ & 81\\
$0.075^b$ & $179(\pm 1)$ & $179(\pm 1)$ & $179(\pm 1)$ &~& $178(\pm 2)$ & 178 \\
\hline
~ &  $\Delta r=0.025$~~ & $\Delta r=0.05 $~~& $\Delta x=0.065$~~ & $\Delta x=0.13 $ \\
\hline
$0.1^c$ &  $15.0(\pm 0.1)$ & $15.0 (\pm 0.1)$ & $15.0(\pm 0.1)$ & $15.0(\pm 0.1)$ 
        & $15.0 (\pm 0.2)$ \\
$0.0875^c$ &  $17.3(\pm 0.2)$ & $17.3(\pm 0.2)$ & $17.3(\pm 0.2)$ & $17.3(\pm 0.2)$ 
        & $17.3 (\pm 0.2)$ \\
$0.075^c$ &  $20.8(\pm 0.3)$ & $20.8(\pm 0.3)$ & $20.8(\pm 0.3)$ & $20.8(\pm 0.3)$ 
        & $20.8 (\pm 0.3)$ \\
$0.0625^c$ &  $24.2(\pm 0.1)$ & $24.2(\pm 0.1)$ & $24.2(\pm 0.1)$ & $24.2(\pm 0.1)$ 
        & $24.2 (\pm 0.3)$ \\
$0.05^c$ &  $30.3(\pm 0.5)$ & $30.3(\pm 0.5)$ & $30.3(\pm 0.5)$ & $30.3(\pm 0.5)$ 
        & $30.3 (\pm 0.4)$\\
\hline
~ &  $\Delta U/C_h \approx 1$~~ & $\Delta U/C_h \approx 2 $~&~&\\
\hline
$2.75^d$ &  $9.31(\pm 0.03)$ & $9.31(\pm 0.03)$ &~&~& $9.31 (\pm 0.02)$ \\
$2.50^d$ &  $9.99(\pm 0.03)$ & $9.98(\pm 0.03)$ &~&~& $9.99 (\pm 0.02)$ \\
$2.25^d$ &  $10.92(\pm 0.03)$ & $10.90(\pm 0.03)$ &~&~& $10.92 (\pm 0.02)$ \\
$2.00^d$ &  $12.58(\pm 0.04)$ & $12.56(\pm 0.04)$ &~&~& $12.58 (\pm 0.02)$ \\
$1.75^d$ &  $15.34(\pm 0.05)$ & $15.31(\pm 0.05)$ &~&~& $15.34 (\pm 0.04)$ \\
\hline
\end{tabular}
\caption{
 MFPTs for (a) Kramer's potential, (b) rough 1 dim. (c) 2-dim. 
entropic barrier and (d) the folding of a polymer chain. The results of the Projective 
Dynamics method are given for their corresponding intervals 
$\Delta \zeta =\Delta l ~~\text{Kramer's,}~ \Delta r ~\text{and}~ \Delta x ~~\text{2-dim, 
and}~~\Delta U/C_h ~~\text{polymer.}$ 
Also shown results obtained by direct measurement 
of the Brownian motion process and the semianalytical solution for the $d$$=$$1$ 
absorption processes.
\label{Tab2a}
}
\end{table*}  
\endgroup

\begin{figure}[ht]
    \includegraphics[width=\linewidth]{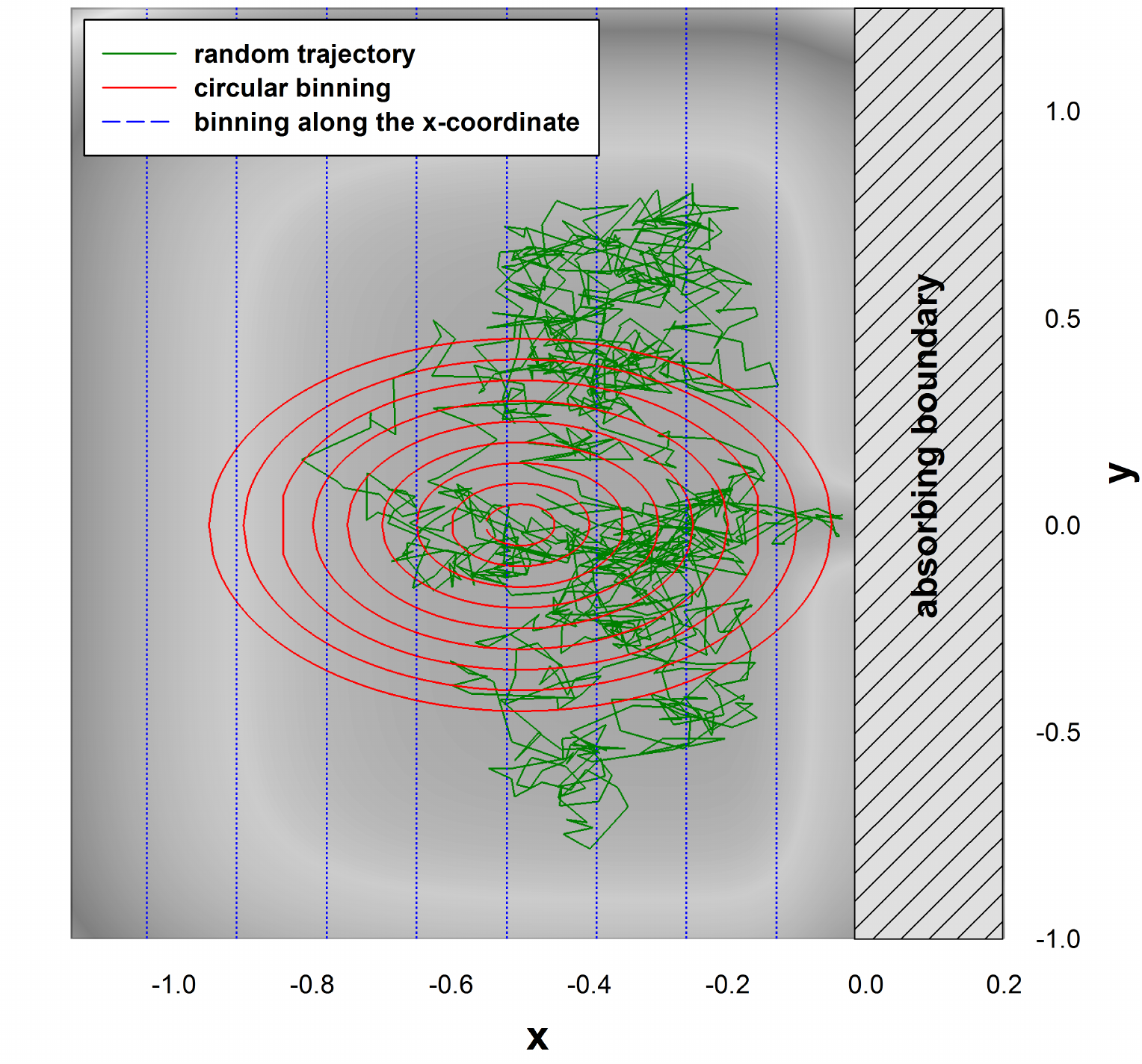}
\caption[]{(color online) The entropic barrier, the states are defined along the coordinate $\zeta=x$ 
(representing a 
progress toward the absorbing boundary) and $\zeta=r$ (representing circular intervals) of some fixed
interval length $\Delta \zeta =\Delta x ~\text{and}~\Delta r$.}
\label{entropic}
\end{figure}

As a $d$$=$$2$ example we applied the PD method to the diffusion 
process at the entropic barrier $U_{entr}(x,y)$ of Ref.~\cite{elber}. 
We chose the coordinate $\zeta$ in two different ways. 
The trajectories are started at $(x_o,y_o)$$=$$(-0.5,0)$ and terminated once they reach $x$$>$$0$. 
We obtained 
results for defining the states both with the coordinate $\zeta$$=$$x$ and $\zeta$$=$$r$ 
[Fig.~(\ref{entropic})]. 
The first represents a binning along the x-coordinate, interpreted as the progress 
towards the absorbing boundary, while 
the circular binning has no such interpretation.  
For $\zeta$$=$$x$ the states were 
$(k-1)\Delta x$$\le$$x$$<$$k\Delta x$, where we chose $\Delta x$$=$$0.065$ and~$0.13$.  
For $\zeta$$=$$r$, the states are defined as
$(S-k)\Delta r$$\le$$r$$< (S-k+1)\Delta r$ for $2\le k\le S$, where $r$ denotes the distance from 
$(x_o,y_o)$.  
We chose $\Delta r$$=$$0.025$ and~$0.05$.
For the circular binning, the bin $k$$=$$1$ is comprised of all points 
with $x$$\le$$0$ outside the circular annulii. 
The results in Table~\ref{Tab2a} showing independence of the MFPT 
illustrate the strength of the PD method.  

As a polymeric example, we applied the PD method to the folding process 
of a model of a small linear polymer chain having the chemical 
structure $(HP_3)_5H$, where $H$ stands for the hydrophobic and $P$ for the polar monomers. 
The structure of the individual chain is 
maintained by three interatomic potentials: a bond-stretching potential, a bond-angle bending 
potential, and a rejecting potential avoiding an overlap of
$HP$ or $PP$ monomers. The chain undergoes a folding process from the fully
elongated state to a compact near native state caused by attractive Lennard-Jones type of 
interactions between $HH$ monomers. 
The PD method is given by choosing the coordinate $\zeta$ as the potential 
energy of the native contacts (the $HH$ interaction energy).
Although this choice seems apparent, it does not correspond to an actual 
dynamical coordinate.  
Starting with the fully elongated chain, for different temperatures of the external bath, 
we simulated the folding process due to the Brownian motion.  
We compared the MFPTs of the chain of reaching a potential energy 
deep enough in the basin of attraction of the energy ground state 
(i.e.\ the native state) so that no escape (i.e.\ unfolding) from there is likely to occur. 
The results of the PD method as well as a direct measurement are shown in Table~\ref{Tab2a}. 
For all simulated temperatures the PD method gives the same results (within statistics) 
as a direct measurement.  

\section{Conclusion and Discussion}
The Projective Dynamics (PD) method, as a tool for projecting multi-dimensional systems onto 
one set of states $\left\{\zeta_k\right\}$ and mapping the time-evolution onto 
a discrete master equation with nearest neighbor coupling, was shown to be 
independent of the
choice in the states $\zeta_k$ and the `length' of the intervals  which 
define the states.
Provided only transitions between 
adjacent states occur,
the PD method correctly obtains MFPTs. This result makes 
the PD method generally 
applicable to the dynamics of any system (discrete or continuous, Markovian or non-Markovian, 
stochastic or deterministic).  
Subject only to the nearest-neighbor coupling constraint, 
the states $\left\{\zeta_k\right\}$ can be chosen in multiple ways, 
without changing the mean first-passage time (MFPT). 
In many cases of interest, the average occupation probability for every state 
$\zeta_k$ is also the same as for the original dynamical system.  
The calculation of the growing and shrinking rates required in the PD method may 
for some models and dynamics be obtained by using other methods, 
such as calculations of histories \cite{Gulb06} or 
forward flux sampling \cite{Allen05} or accelerated molecular dynamics \cite{Miron04,Voter02}.  

\begin{acknowledgments}
Useful discussions with Jerzy Blawzdziewicz, Weinan E, Ron Elber, Miroslav Kolesik, 
Cory O'Hern, Per Arne Rikvold, and Eric Vanden-Eijnden are gratefully acknowledged, 
as well as useful interactions at a 2008 summer school at the Aspen Center for Physics.  
This research was supported in part by the National 
Science Foundation through TeraGrid resources provided by the 
Texas Advanced Computing Center (TACC) under grant 
number TG-DMR090092.  
\end{acknowledgments}

\end{document}